\documentclass[prb,
superscriptaddress,showpacs,amsmath,amssymb]{revtex4}

\begin{document}

\title{Thermodynamics of Kerr black hole: Tsallis-Cirto composition law and entropy quantization}

\author{G.E.~Volovik}
\affiliation{Landau Institute for Theoretical Physics, acad. Semyonov av., 1a, 142432,
Chernogolovka, Russia}

\date{\today}

\begin{abstract}
 The processes of splitting and merging of black holes obey the composition law generated by the Tsallis-Cirto $\delta=2$ statistics. The same composition law expresses the
 full entropy of the Reissner-Nordstr\"om black hole via the entropies of its outer and inner horizons. Here we apply this composition law to the thermodynamics of the Kerr black hole.  As distinct from Reissner-Nordstr\"om black hole, where the full entropy depends only on mass $M$ and does not depend on its charge $Q$, the entropy of Kerr black hole is the sum of contributions from its mass $M$ and angular momentum $J$, i.e. $S(M,J)=S(M,0) + 4\pi \sqrt{J(J+1)}$. Here $S(M,0)$ is the entropy of the Schwarzschild black hole. This demonstrates that when the Kerr black hole with $J\gg 1$ absorbs or emits a massless particle with spin $s_z=\pm 1/2$, its entropy changes by $|\Delta S| = 2\pi$. We also considered the quantization of entropy suggested by the toy model, in which the black hole thermodynamics is represented by the ensemble of the Planck-scale black holes -- Planckons.
 The Tsallis-Cirto composition law is also extended to the thermodynamics of Kerr-Newman black hole.
  \end{abstract}
\pacs{
}

\maketitle

\tableofcontents

\section{Introduction}

Hawking radiation from and Bekenstein entropy of the black holes with multiple horizons and the corresponding thermodynamics can be different from that which is obtained by consideration of only the effects of the outer horizon, see e.g. the recent paper\cite{Singha2025} and references therein. 
For the spherically symmetric Reissner-Nordstr\"om (RN) black hole with two horizons, several different approaches have been used.\cite{Volovik2021,Volovik2025TC} This includes: (i) coherent Hawking radiations from two horizons; (ii) macroscopic quantum tunneling; (iii) adiabatic transformations; (iv)  singular coordinate transformations; and also (v) the Tsallis-Cirto non-extensive statistics.\cite{TsallisCirto2013,Tsallis2020} All these approaches demonstrate that the total entropy of the black hole is not determined by the area of the outer horizon. Due to the coherent correlations between the outer and inner horizons, the total entropy of the RN black hole and the temperature of Hawking radiation depend only on mass of the black hole and do not depend on the black hole charge. 

In the paper \cite{Volovik2021}, an attempt was made to consider the entropy of rotating black holes -- the Kerr black hole and the Kerr-Newman black hole. The method of adiabatic transformations was used for this purpose. However, it turned out that this method is inapplicable for the rotating case. Here, we consider the thermodynamics of rotating black holes using non-extensive Tsallis-Cirto statistics.
In Sec. \ref{TCentropy} we recall the Tsallis-Cirto type statistics, which provides a composition rule for black hole merger and splitting processes, as well as for the combination of the entropies of the inner and outer horizons of an RN black hole. This composition law is applied to 
Kerr and Kerr-Newman black holes in Sec. \ref{rotating}.

\section{Composition rule and Tsallis-Cirto $\delta=2$ entropy}
\label{TCentropy}

\subsection{Composition rule for merging and splitting of black holes}

The black hole entropy $S_{\rm BH}(M)=4\pi GM^2$ is non-extensive with the special type of composition.  One example is the process of the splitting of the black hole with mass $M$ into two smaller black holes with masses $M_1$ and $M_2$. For $M_1+M_2=M$, the entropy obeys the following composition rule:
\begin{equation}
\sqrt{S_{\rm BH}(M=M_1 +M_2)}= \sqrt{S_{\rm BH}(M_1)} +\sqrt{S_{\rm BH}(M_2)}\,.
\label{BlackHoles}
\end{equation}
This composition suggests  the application of the non-extensive Tsallis-Cirto $\delta=2$ entropy:\cite{TsallisCirto2013,Tsallis2020}
\begin{equation}
S_{\delta =2}=\sum_i p_i \left(\ln\frac{1}{p_i} \right)^2\,,
\label{TCentropy}
\end{equation}
which gives for a system composed of two probabilistically independent subsystems $A$ and $B$, the following non-additive composition rule:
\begin{equation}
\sqrt{S_{\delta =2}(A+B)}=\sqrt{S_{\delta =2}(A)} + \sqrt{S_{\delta =2}(B)}\,.
\label{TCentropy2}
\end{equation}

It was shown\cite{Volovik2025TC} that this composition law can be extended to the white hole horizons with negative entropy. For example, if the processes of splitting and merging include both the black holes with positive entropy and the white holes with negative entropy, the Eq.(\ref{BlackHoles}) becomes
\begin{equation}
\sqrt{|S_{\rm BH}(M=M_1 +M_2)|}= \sqrt{|S_{\rm BH}(M_1)|} +\sqrt{|S_{\rm BH}(M_2)|}\,.
\label{BlackWhiteHoles}
\end{equation}

\subsection{Composition rule for inner and outer horizons of Reissner-Nordstr\"om black hole}

 Another example of the application of the Tsallis-Cirto non-extensive statistics is provided by the multi-horizon black holes with positive and negative entropies of horizons. For the Reissner-Nordstr\"om (RN) black hole, the composition law expresses the full entropy of the RN black holes in terms of the entropies of its outer and inner horizons:\cite{Volovik2025TC}
\begin{equation} 
\sqrt{S_{\rm RN}}= \sqrt{S_+} +\sqrt{|S_-|}\,.
\label{RNBL}
\end{equation}

The entropies of the outer and inner horizon are
\begin{equation}
S_{\rm RN}(r_+)= \pi r_+^2/G=\pi G \left(M+ \sqrt{M^2 -\alpha Q^2/G} \right)^2\,,
\label{outerRN}
\end{equation}
 \begin{equation}
S_{\rm RN}(r_-)= -\pi r_-^2/G=-\pi G \left(M- \sqrt{M^2 -\alpha Q^2/G} \right)^2\,.
\label{innerRN}
\end{equation}
Here $\alpha$ is the fine structure constant (actually the running coupling) and $Q$ is the integer valued electric charge with $Q=-1$ for electron. The minus sign in Eq.(\ref{innerRN}) is because the inner horizon is the white horizon with negative entropy (in the Arnowitt-Deser-Misner formalism, \cite{ADM2008} the shift function ${\bf v} $ is directed towards the black horizon, while it has the opposite direction in case of the white horizon).

Then from Eq.(\ref{RNBL}) it follows that the total entropy of the RN black hole is
 \begin{equation}
S_{\rm RN}(M)= \left( \sqrt{S_{\rm RN}(r_+)} + \sqrt{|S_{\rm RN}(r_-)|}\right)^2= \pi(r_+ +r_-)^2/G =4\pi GM^2\,.
\label{entropyRN}
\end{equation}
It does not depend on the electric charge and is fully determined by the mass $M$.
There is a natural explanation for this: when the parameter $\alpha$ changes adiabatically to zero at a fixed $M$, the entropy does not change, and thus it is the same as the entropy of the Schwarzschild black hole with the same mass $M$.

In our earlier paper,\cite{Volovik2021} it was suggested that the similar situation may happen for the rotating black hole, where the adiabatic change of the parameter $\hbar$ was considered and it was concluded that the entropy of the Kerr black hole is only determined by its mass.
However, this is not so, see the next Section \ref{rotating}. The reason why this conclusion is incorrect is that changing the parameter $\hbar$ at a fixed angular momentum $L=\hbar J$ is not an adiabatic process, since in this process the quantum number $J$ experiences discrete quantum jumps.

\section{Rotating black holes}
\label{rotating}

\subsection{Entropy of Kerr black hole from composition law}
\label{KerrSec}

Let us apply the non-extensive Tsallis-Cirto  $\delta=2$ statistics\cite{TsallisCirto2013,Tsallis2020} to the composition law for the inner and outer horizons of the Kerr black hole. According to this composition law the total entropy of the Kerr black hole is expressed in terms of the entropies of the inner and outer horizons in the same way as for the RN black hole in Eq.(\ref{entropyRN}), i.e.:
\begin{equation} 
\sqrt{S_{\rm Kerr}}= \sqrt{S_+} +\sqrt{|S_-|}\,.
\label{BHTsallis}
\end{equation}
Here for the outer horizon we have
\begin{equation}
S_+ = 2\pi GM^2 \left(1+\sqrt{1-J^2/M^4G^2}   \right)\,,
\label{outer}
\end{equation}
and the entropy of the inner horizon is negative with:
\begin{equation}
|S_-| = 2\pi GM^2 \left(1- \sqrt{1-J^2/M^4G^2}   \right)\,.
\label{inner}
\end{equation}
 
Then the composition law in Eq.(\ref{BHTsallis}) gives the following total entropy of the Kerr black hole, which contains two separate contributions from the mass $M$ and from rotation with angular momentum $J$:
\begin{equation}
S_{\rm Kerr}(M,J)= 4\pi GM^2 + 4\pi J=S_{\rm Schwarzschild}(M) + 4\pi J\,.
\label{KerrEntropy}
\end{equation}

\subsection{Entropy quantization for Kerr black hole}
\label{KerrQuantSec}

Eq.(\ref{KerrEntropy}) suggests the quantization of the rotational part of the total entropy:
\begin{equation}
S_{\rm Kerr}(M,J)-S(M,0) =  4\pi \sqrt{J(J+1)}\,.
\label{quantization}
\end{equation}
In the thermodynamic limit $J\gg 1$ this gives
\begin{equation}
{\partial S}/{\partial J}{\big |}_M =  4\pi\,.
\label{dSdJ}
\end{equation}

If the Kerr black hole with $J\gg 1$ absorbs or emits a massless particle with spin $s_z=\pm 1/2$,
then the mass $M$ of the black hole remains the same, while its entropy changes by the following amount:
\begin{equation}
|\Delta S| = 2\pi\,.
\label{EntropyChange}
\end{equation}
Such stepwise behaviour of the black hole entropy is discussed in many papers, starting with Bekenstein who argued that entropy should be quantized in equidistant steps,\cite{Bekenstein1973,Bekenstein1974} see e.g. the recent paper\cite{Jana2025} and references therein.
But in Eq.(\ref{quantization}) only the rotational part of the entropy of RN black hole is quantized.

The entropy of the extremal Kerr black hole with $J=GM^2$ is
\begin{equation}
S_{\rm Kerr \, extreme}=S(M, J=GM^2) =2S(M, 0)= 8\pi J\,.
\label{KerrExtremal}
\end{equation} 
It is four times the traditionally discussed value $S_0=2\pi J$. 
Also, if Eq.(\ref{KerrExtremal}) is written in the form
\begin{equation}
S_{\rm Kerr\,extreme}= 8\pi \sqrt{J(J+1)}\,,
\label{KerrExtremal2}
\end{equation} 
  this corresponds to Eq.(11) in Ref. \cite{Khriplovich1998}  with an additional factor of 4. Instead of being two times smaller than the entropy of a Schwarzschild black hole of the same mass $M$, it is two times larger.

\subsection{Entropy of Kerr-Newman black hole from the composition law}
\label{KerrNewmanSec}

Let us apply the composition rule in Eqs. (\ref{RNBL}) and (\ref{BHTsallis}) to the more complicated black hole  -- the Kerr-Newman black hole with angular momentum $J$ and charge $Q$. From the entropies of inner and outer horizons of the Kerr-Newman black hole one obtains:
\begin{equation}
S(M,J,Q)= S(M,0,0) + 2\pi\left( \sqrt{\alpha^2 Q^4 +4J^2} - \alpha Q^2 \right).
\label{KerrNewman}
\end{equation} 
For $Q=0$ this transforms into the equation (\ref{KerrEntropy}) for the entropy of the Kerr black hole, and for $J=0$ this transforms into the equation (\ref{entropyRN}) for the entropy of the RN black hole. 

The entropy of Kerr-Newman black hole depends on the running coupling $\alpha$, but this dependence disappears in the RN limit, i.e. at $J=0$, where the entropy depends only on mass $M$. The reason is that at $J=0$ we can adiabatically transform the RN black hole by slow change of this parameter $\alpha$ to zero at fixed mass $M$ and fixed quantum number $Q$. Since in the adiabatic process the entropy of the RN black hole does not change, it is the same as for the Schwarzschild black hole with the same mass $M$.\cite{Volovik2021} At $J\neq 0$ such an adiabatic process is absent, since there are too many parameters that need to be fixed.

\subsection{Entropy quantization in Planckon toy model}
\label{PlanckonSec}

Planckon model is the toy model, in which the black hole is considered as an ensemble of Planck-scale quanta, Planckons.\cite{Volovik2025P} 
The Tsallis-Cirto $\delta=2$ statistics with the non-additive composition law in Eq.(\ref{TCentropy2}) suggests that the entropy of the ensemble is determined by the   correlations between Planckons. It is proportional to the number of the Planckon pairs, i.e. the number of ways to select two Planckons from an ensemble of $N$ Planckons:
\begin{equation}
S_{\rm BH}(N) =C \begin{pmatrix}N
\\ 2\end{pmatrix} =C\frac{N! }{2! (N-2)!}=C\frac{N(N-1)}{2} \,.
\label{BWHolesN2}
\end{equation}
Here the dimensionless parameter $C$ depends on the choice of the Planckon mass quantum.
 In Ref. \cite{Volovik2025P} the mass of Planckon was chosen as the reduced Planck mass, $m_{\rm P}=1/\sqrt{8\pi G}$, which corresponds to $C=1$. If we choose the Planck mass $M_{\rm P}=1/\sqrt{G}$ as the mass quantum, i.e. $M= N M_{\rm P}$, one has $C=8\pi$.
 
 All this suggests that for the Kerr black hole and with $M= N M_{\rm P}$ one would have:
\begin{equation}
S_{\rm Kerr}(M,J)\equiv S_{\rm Kerr}(N,J)=4\pi \left(N( N-1)+  \sqrt{J(J+1)}\right)\,.
\label{KerrPlanck2}
\end{equation}
In this toy model the entropy depends on two quantum numbers, $N$ and $J$, with $J=N^2$ for the extremal RN black hole in the semiclassical limit $N\gg 1$. Unlike quantization by angular momentum $J$, quantization by Planckon number $N$ is the property of a toy model. The factor $4\pi$ takes place only for the Planckon mass quantum $M_{\rm P}=1/\sqrt{G}$. Nevertheless, for this mass quantum, the extreme regime is expressed entirely in integers, $J=N^2$, which is quite intriguing. 

Note that this toy model produces the results, which are similar to that of the $SU(N)$ matrix model discussed in Refs.\cite{Chu2025,Chu2025b}, where the horizon of Schwarzschild black hole is identified with the fuzzy sphere and $N$ is the dimension of the Hilbert space of quantum gravity. Also, the $N^2$ entropy in terms of the $N\times N$ Hermitian matrices was considered for the de Sitter cosmological horizon -- the dS-Matrix Theory.\cite{Susskind2023}

 For the charge black holes, the quantization is violated by the parameter $\alpha$ -- the running coupling. For example, the running coupling $\alpha$ in Eq.(\ref{outerRN}) enters the condition for extremality, $N=\sqrt{\alpha} Q$, and thus $N$ cannot be integer. The running coupling $\alpha$ also enters the entropy of the Kerr-Newman black hole in Eq.(\ref{KerrNewman}).

\subsection{Thermodynamics of rotating black holes}

The temperature and entropy of the RN black hole with two horizons  are determined by the correlations between the inner and outer horizons. This gives rise to the thermodynamics of the RN black hole, which differs from the traditional thermodynamics determined only by the outer horizon.\cite{Davis1977} The same takes place for the Kerr black hole. Using  Eq.(\ref{KerrEntropy}) and thermodynamic relation
\begin{equation}
dM=TdS+{ \boldsymbol\Omega}\cdot d{\bf J} \,,
\label{ThermoEq}
\end{equation}
one obtains  temperature of the rotating black hole:
\begin{equation}
T=\frac{\partial M}{\partial S} \bigg|_{\bf J} =\frac{1}{8\pi GM}\,.
\label{Temperature}
\end{equation}
Similar to the RN black hole, this temperature does not deviate from the temperature of the Schwarzschild black hole with the same mass.

Eq.(\ref{KerrEntropy}) and thermodynamic relation Eq.(\ref{ThermoEq}) also determine the angular velocity of the Kerr black hole:
\begin{equation}
{\boldsymbol\Omega}=\frac{\partial M}{\partial{\bf J}} \bigg|_S=- \frac{\bf J}{2GMJ}  \,.
\label{AngularVelocity}
\end{equation}
The angular velocity demonstrates the negative moment of inertia and it has the following connection with the temperature in Eq.(\ref{Temperature}), see also Eq.(\ref{dSdJ}):
\begin{equation}
\Omega\equiv |{\boldsymbol\Omega}|=4\pi T \,.
\label{OmegaTConnection}
\end{equation}

For the Kerr-Newman black hole with entropy in Eq.(\ref{KerrNewman}) one has:
\begin{equation}
T=\frac{1}{8\pi GM}\,\,,\,\, {\boldsymbol\Omega}=-\frac{ 8\pi T }{\sqrt{4J^2 
+\alpha^2Q^4}} \,{\bf J}\,.
\label{OmegaTNewman}
\end{equation}
It should be noted that there is no discontinuity in heat capacity here, unlike the thermodynamics of rotating black hole based on the outer horizon.\cite{Davis1977}

The connection between $\Omega$ and $T$ suggests the possible modification of the Zel'dovich-Starobinsky rotational superradiance.\cite{Zeldovich1971,Zeldovich1972,Starobinsky1973,StarobinskyChurilov1974}
At external temperatures, spontaneous emission can occur even at $\omega > m\Omega$.\cite{Endlich2017} The black hole temperature in the equation (\ref{OmegaTConnection}) can play a similar role.

\section{Composition law for Schwarzschild-de Sitter black hole}
\label{SdS}

It appears that the non-additive Tsallis-Cirto $\delta=2$ statistics with the composition law in Eq.(\ref{TCentropy2}) plays the important role in the  determination of the entropy of the compact object -- black hole -- in Minkowski environment. If so, can this be applied to a black hole in the de Sitter environment with its cosmological horizon? In this regard, the Schwarzschild-de Sitter (SdS) black hole was considered by Susskind in his dS Matrix Theory. A discrepancy of a factor of 2 between the composition law in equation (\ref{TCentropy2}) and the calculated entropy deficit was found.\cite{Susskind2023}
Let's try to correct this approach.

Here we consider the composition law for the SdS entropy using approach based on the singular coordinate transformations. This approach was used in particular for the calculations of the probability of the macroscopic quantum tunneling from black hole to white hole.\cite{Volovik2022} Note that if the coordinate transformation has singularity, the diffeomorphism invariance is not necessary applicable, and the coordinate transformation may connect different physical objects. A particular example is
 the coordinate transformation from the black hole with mass $M$ to the white hole with the same mass, $dt\rightarrow dt+ 2dr \frac{v}{1-v^2}$, where $v^2=2GM/r$. This transformation has singularity at the horizon, i.e. at $v^2(r)=1$.
 
Now let us consider the SdS black hole with $v^2(r)=H^2r^2 +\frac{2MG}{r}$, which has two horizons: the black hole horizon at $r=r_-$ and the cosmological horizon at $r=r_+>r_-$. Let us  make the singular coordinate transformation $r\rightarrow -r$. Then one obtains the state with  $v^2(r)=H^2r^2 - \frac{2MG}{r}$, which corresponds to the negative mass of black hole. This state has only a single horizon at $r_0=r_+ + r_-$. The entropies of the initial and final states obey the composition law:
\begin{equation}
\sqrt{S(r_0)}= \sqrt{S(r_-)} +\sqrt{S(r_+)}\,.
\label{SdScomposition}
\end{equation}
This corresponds to the Tsallis-Cirto $\delta=2$ statistics and thus justifies the dS Matrix Theory suggested by Susskind.\cite{Susskind2023}

Whether negative-mass objects exist in the sky is an open question.\cite{Odintsov2026}
But this does not affect the consideration of the entropy composition law in Eq.(\ref{TCentropy2}).
The same composition law has been suggested for the entanglement/correlation entropy between the geometric and dual sectors of doubled spacetime.\cite{Chouha2026}

\section{Conclusion}
\label{ConclusionSec}

We applied the composition law generated by the Tsallis-Cirto $\delta=2$ statistics to the Kerr black hole and expressed the entropy of black hole via the entropies of its outer and inner horizons.  In this approach we obtained that the entropy of Kerr black hole is the sum of contributions from its mass $M$ and angular momentum $J$, i.e. $S(M,J)=S(M,0) + 4\pi \sqrt{J(J+1)}$, where $S(M,0)$ is the entropy of the Schwarzschild black hole. In the thermodynamic limit $J\gg 1$ this gives
${\partial S}/{\partial J}{\big |}_M =  4\pi$, while the angular velocity of the rotating black hole is expressed in terms of its temperature, $\Omega=4\pi T$. When the Kerr black hole with $J\gg 1$ absorbs or emits a massless particle with spin $s_z=\pm 1/2$, its entropy changes by $|\Delta S| = 2\pi$.

This is different from the thermodynamics of the Reissner-Nordstr\"om black hole, where the application of the Tsallis-Cirto $\delta=2$ statistics shows that the full entropy of the RN black hole depends only on mass $M$ and does not depend on its charge $Q$, i.e. $S(M,Q)=S(M,0)$. This agrees with possibility to adiabatically transform the entropy of the RN black hole to the entropy of the Schwarzschild black hole by adiabatically reducing the fine structure constant $\alpha$ to zero.

Using the Tsallis-Cirto $\delta=2$ statistics we also obtained the entropy of the Kerr-Newman black hole with its inner and outer horizons, $S(M,J,Q)= S(M,0,0) + 2\pi\left( \sqrt{\alpha^2 Q^4 +4J^2} - \alpha Q^2 \right)$. As distinct from the RN black hole with $J=0$, this entropy does depend on $\alpha$ parameter. This demonstrates that while there is an adiabatic path between the Reissner-Nordström and Schwarzschild black holes, there is no adiabatic path between the Kerr-Newman and Kerr black holes.

To further substantiate the applicability of the Tsallis-Cirto $\delta=2$ statistics to the composition law for event horizons, we considered a Schwarzschild-de Sitter black hole and demonstrated that the dS Matrix Theory proposed by Susskind\cite{Susskind2023} indeed works.

\end{document}